\documentclass[aps,prl,twocolumn,showpacs,superscriptaddress]{revtex4}

\usepackage{hyperref}
\usepackage{amssymb}
\usepackage{amsmath}
\usepackage{epsfig}

\begin{document}
\title{Quantum interference of Glory rescattering in strong-field atomic ionization}

\author{Q. Z. Xia}
\affiliation{National Laboratory of Science and Technology on Computational Physics, Institute of Applied Physics and Computational Mathematics, Beijing 100088, China}
\author{J. F. Tao}
\affiliation{National Laboratory of Science and Technology on Computational Physics, Institute of Applied Physics and Computational Mathematics, Beijing 100088, China}
\author{J. Cai}
\affiliation{School of Physics and Electronic Engineering, Jiangsu Normal University, Xuzhou 221116, China}
\author{L. B. Fu}
\affiliation{Graduate School, China Academy of Engineering Physics, Beijing 100088, China}
\affiliation{CAPT, HEDPS, and IFSA Collaborative Innovation Center of MoE, Peking University, Beijing 100871, China}
\author{J. Liu}
\email[]{liu\_jie@iapcm.ac.cn}
\affiliation{National Laboratory of Science and Technology on Computational Physics, Institute of Applied Physics and Computational Mathematics, Beijing 100088, China}
\affiliation{CAPT, HEDPS, and IFSA Collaborative Innovation Center of MoE, Peking University, Beijing 100871, China}

\begin{abstract}
{
During the ionization of atoms irradiated by linearly polarized intense laser fields, we find for the first time that the transverse momentum distribution of photoelectrons can be well fitted by a squared zeroth-order Bessel function because of the quantum interference effect of Glory  rescattering. The characteristic of the Bessel function is determined by the common angular momentum of a bunch of semiclassical paths termed as Glory  trajectories, which are launched with different nonzero initial transverse momenta distributed on a specific circle in the momentum plane and finally deflected to the same asymptotic momentum, which is along the polarization direction, through post-tunneling rescattering. Glory rescattering theory (GRT) based on the semiclassical path-integral formalism is developed to address this effect quantitatively. Our theory can resolve the long-standing discrepancies between existing theories and experiments on the fringe location, predict the sudden transition of the fringe structure in holographic patterns, and shed light on the quantum interference aspects of low-energy structures in strong-field atomic ionization.
}
\end{abstract}

\maketitle

\paragraph{Introduction.---}
As a beautiful phenomenon, optical Glory  is a series of bright concentric rings that surround the observer's shadow when light is backward scattered\cite{book_Nussenzveig,Adam2002}.
In 1959, its semiclassical counterpart in quantum scattering\cite{wheeler} was identified and associated with a specific singularity, i.e., the axial caustic singularity \cite{Berry2015}.
 Similar to other singularities ranging from critical phenomena to black holes, Glory scattering has been explored
and is expected to be a good probe of physical processes in a number of areas, such as nuclear physics, atomic physics, and gravitation\cite{Poelsema1982, Hussein1984, Ueda1998, Roberts2002, AngewChem2005,Morette1984}. In this Letter, we report the emergence of the Glory effect in strong-field atomic ionization.

As the fingerprint of the Glory effect, we find that the transverse photoelectron momentum distribution in
atomic ionization is well fitted by the square of a zeroth-order Bessel function.
  This finding can be explained by the quantum interference of an infinite number of Glory  trajectories (GTs), which are launched with nonzero initial transverse momenta distributed on a specific circle in the momentum plane and finally deflected to the same asymptotic momentum along the polarization direction by post-tunneling forward rescattering. The axial caustic singularity associated with GTs could lead to the breakdown of the traditional two-path  quantum interference scenario in strong-field ionization dynamics.

A nonperturbative Glory rescattering theory (GRT) is developed in this Letter, which self-consistently includes the Coulomb-laser coupling\cite{Smirnova2007} within the framework of  the coordinate configuration path-integral representation \cite{Kleinert}.
Our theory can provide insight into the Glory effect by resolving the infinite co-dimension caustic structure \cite{Berry1976}, i.e., Glory caustic, in strong-field rescattering.
Using GRT, we can successfully resolve the discrepancies between existing theories based on two-path interference scenario and experiments
on the fringe location \cite{Huismans,huismansPRL}, and predict a sudden transition of the fringe structure in the holographic pattern\cite{Hickstein_2012} of strong-field atomic ionization. Its implications  in the low-energy  spectrum\cite{Blaga_2009, Quan_2009} of photoelectrons are also discussed.

\begin{figure}[t]
\includegraphics[width=1.0\linewidth]{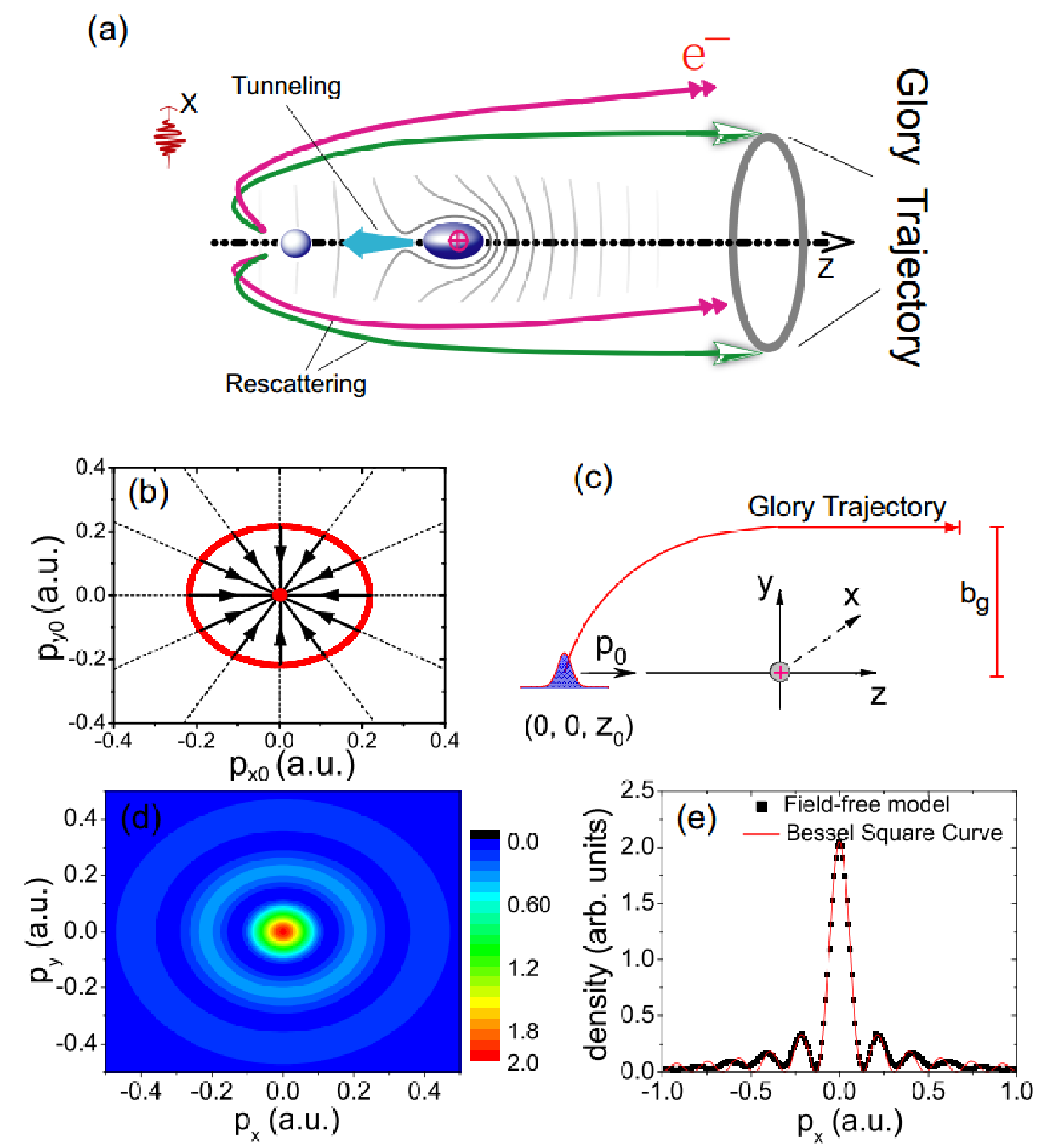}
\caption{(Color online) (a)Glory  scattering during the forward rescattering after electron tunnelling.
Traditional two-path (magenta) interference and  Glory interference of an infinite number of paths (green trajectory and its rotational counterparts) that approach the same final momentum.  For details, refer to the text.
(b)Axial caustic singularity in Glory scattering: GTs projected on the transverse momentum plane, which are launched from the red circle and finally converge at the origin. (c) Illustration of  Glory scattering of localized wavepacket.
Transverse momentum distribution after scattering in laser-field-free model of 2D (d) and 1D (e). The incident wavepacket is $\psi_0=\exp(-|\vec{r}-\vec{r}_0|^2+i\vec{p}_0\cdot\vec{r})$, where $\vec{r}_0=(0, 0, -40)$ and $\vec{p}_0=(0, 0, 0.5)$. The 2D slice is obtained at $p_z=0.5$, while the 1D (black dot) is obtained at $p_z=0.5$ and $p_y=0$. The curve of $J_0^2(b_gp_x)$ is plotted with a red solid line in (e) for comparison. }
\end{figure}

\paragraph{Axial caustic singularity and Glory effect in rescattering.---}
In the semiclassical description, the photoelectron experiences a three-step process in the Coulomb-laser field, as shown in Fig.\ 1(a). Initially, the electron tunnels out of the binding potential distorted by a strong electric field at time $t_0$. After accelerating under the influence of the linearly polarized (LP) laser field in the second step, the electron can be driven back and rescattered by its parent ion.
The tunneled electron has an initial transverse momentum of $\vec{p}_{\perp0}=(p_{x0},p_{y0})$
\cite{book_JieLiu}. Through the rescattering \cite{recollision}, the electron finally approaches an asymptotic momentum of $\vec{p}_{\perp f}$. In the mapping between $\vec{p}_{\perp0}$ and $\vec{p}_{\perp f}$, a special singularity structure known as an axial caustic singularity emerges. The underlying physical picture is shown in Fig.\ 1(b), where the trajectories launched with initial momenta distributed on a specific circle\cite{Tau_2017} in the momentum plane can finally converge at the origin. The Jacobian $|\partial \vec{p}_{\perp f}/\partial \vec{p}_{\perp 0}|$ will vanish. The corresponding trajectories are named GTs, as plotted in Fig.\ 1(a).

In the traditional Glory scattering,
the classical differential cross section takes the form of $ b(db/d\theta)/\sin{\theta}$ ($b$ is the impact parameter corresponding to the scattering angle $\theta$), whereas in quantum semiclassical theory, the divergent term $1/\sin{\theta}$ is replaced by $2\pi l_g J_0^2(l_g \sin\theta)$ \cite{wheeler, Morette1984}. Here, $l_g$ and $b_g$ are the common angular momentum and impact parameter of the corresponding GTs, respectively. This indicates that GTs dominate the quantum interference and lead to a Bessel-type oscillation. Herein, atomic units are used unless otherwise specified.

 The Glory effect can be illustrated by setting a wavepacket that is scattered by a Coulomb field.  The geometric configuration of  our model calculation is shown in Fig.\ 1(c), in which  a Gaussian wavepacket originates at $\vec{r}_0$
with an average momentum of $\vec{p}_0$\cite{Meckel_2014}.
For the GTs whose asymptotic momenta are along the z axis, their emergent impact parameter $b_g$ can be fixed through the relation $b_g=l_g/p_0=\sqrt{2r_0}/p_0$ by solving the classical Kepler problem. The results of Glory scattering are  presented in panel (d), which  shows a bright spot in the central region surrounded by a series of concentric rings in the transverse momentum plane. In particular, the distribution of $p_x$ is highly consistent with the expression of  $J^2_0(b_g p_x)$ with $b_g p_x\sim l_g\theta$, as shown in panel (e). In contrast, in the well-known Rutherford scattering, a singularity  of type $\sim \sin^{-4}(\theta/2)$ emerges in the expression of the scattering section both classically and in quantum theory\cite{landau}.

 A similar singularity also emerges in the rescattering of the tunneled electron in the combined Coulomb potential and laser field, as illustrated in Fig. 1(a). To address this effect clearly, we develop the GRT based on the semiclassical path-integral formalism in coordinate configuration space\cite{Kleinert}. We find  that in strong-field ionization, near the singular point, the contribution from those GTs dominates the transition amplitude $M_{\vec{p}}=-i\int dt \langle\vec{p}|U(t_f,t)V_L(t)U_0(t,0)|\psi_i\rangle$ of quantum scattering, where U and $U_0$ denote the complete and laser field-free evolution operators, respectively, and $V_L(t)$ denotes the interaction with the laser field. With this recognition,    $M_{\vec{p}}$ can be reduced into the following simple expression after a lengthy deduction\cite{Supp}:
\begin{eqnarray}
|M_{\vec{p}}|^2\sim \varpi P_{\bot g} b_g J_{0}^2(p_\perp b_g).
\label{7_5_4}
\end{eqnarray}
Here, $P_{\bot g}$ is the initial transverse momentum of the GT at the tunneling exit,
$b_g$ is the emergent impact parameter of the GT, and
$\varpi$ is the weight of the  GT based on the initial phase and initial transverse momentum  through the ADK tunneling formula\cite{ADK}.

The corresponding GT can be traced by solving the Newtonian equations of the Hamiltonian $H=\frac{1}{2}(\vec{P}+\vec{A}(t))^2-1/r$ that governs the motions of the rescattered electron in the combined Coulomb potential and laser field of gauge potential $\vec{A}(t)$. Because of the cylindrical symmetry, we can  restrict electron motion on the $x-z$ plane, i.e.,
\begin{eqnarray}
\dot{x}&=&P_x,\ \ \ \ \dot{z}=P_z+A_z; \\
\dot{P}_x&=&-\frac{x}{(x^2+z^2)^{3/2}},\ \ \ \ \ \dot{P}_z=-\frac{z}{(x^2+z^2)^{3/2}}.
\end{eqnarray}
Then, the initial conditions are set as $\omega t=\omega t_0$, $x_0=0$ and $P_{x0}=P_{\bot g}$. The initial coordinate of the tunnel exit, i.e., $z_0$, can be calculated from $\frac{1}{8 z_0}+\frac{1}{16 z_0^2}+\frac{1}{4}\epsilon \cos(\omega t_{0})z_0=\frac{I_p}{4}$\cite{book_JieLiu}.
 In the nonadiabatic setting\cite{ccsfa, LiM_2016}, $P_{z0}=\frac{\epsilon}{\omega}\sin(\omega t_0)\sqrt{1+\gamma(t_0)^2}$, where $\gamma(t_0)=\omega\sqrt{2I_p+P_{x0}^2}/|\epsilon\cos(\omega t_0)|$.

By solving the above equations, we consider the asymptotic condition for  the GT that $(P_x, P_z) \mathop{\rightarrow}\limits_{t\rightarrow \infty} (0, p_{\|})$ and denote $x\mathop{\rightarrow}\limits_{t\rightarrow \infty} b_g$ as the emergent impact parameter. We can then obtain the $P_{\bot g}$ and $b_g$ as a function of the asymptotic momentum  $p_{\|}$ implicitly for the given laser parameters and atomic ionization potential. Then, the transition amplitude formula (1) provides the probability of the  asymptotic momentum $(p_\bot, p_\|)$ of the ionized electrons.

\begin{figure}[t]
\includegraphics[width=1.0\linewidth]{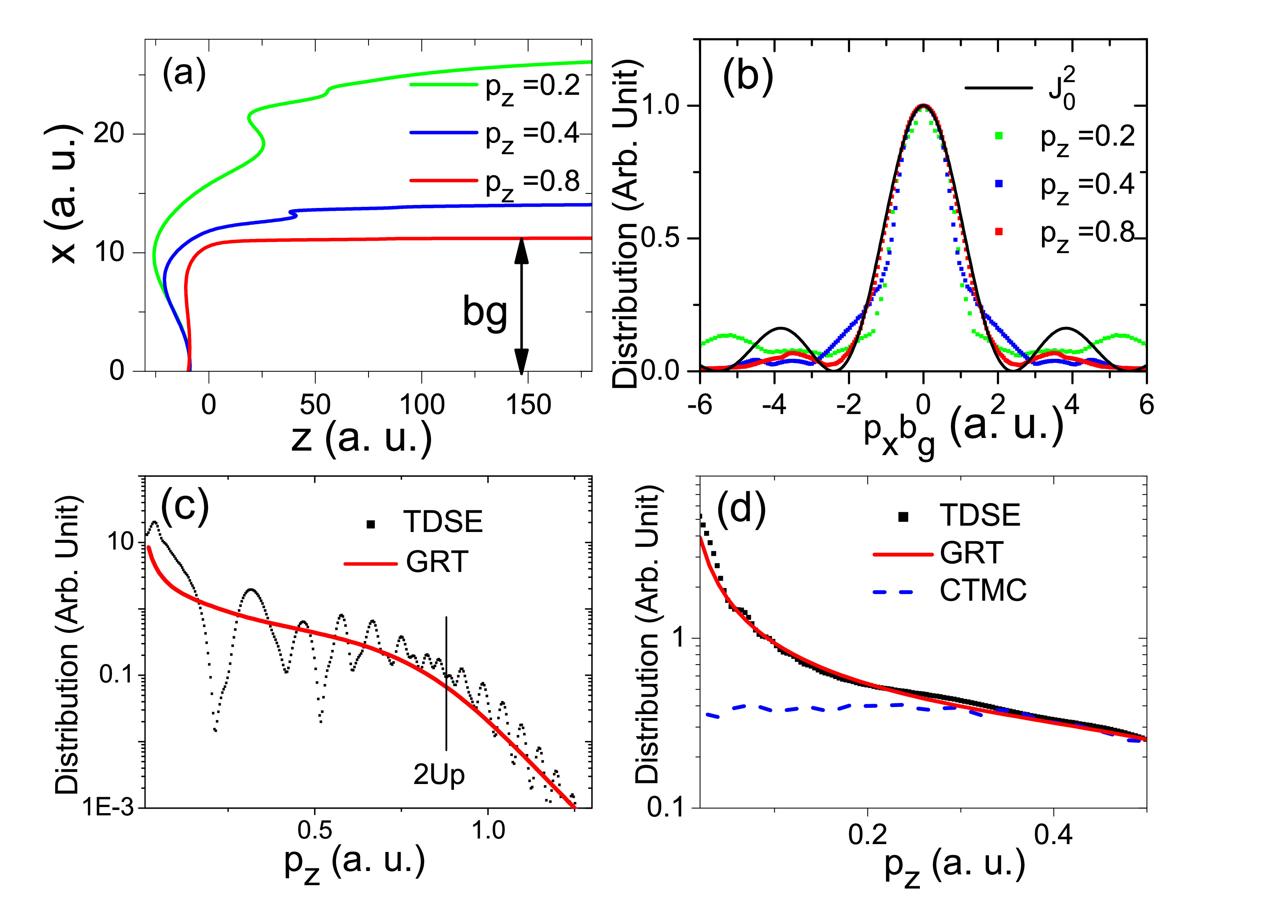}
\caption{(Color online)
(a) Glory  trajectories corresponding to $p_z=0.2, 0.4$, and $0.8$ from top to bottom. (b) Transverse momentum distribution corresponding to $p_z=0.2, 0.4$ and $0.8$.
The squared Bessel function is plotted for comparison.
(c) Simulated result (black squares)  of the longitudinal momentum distribution and the prediction of GRT (red curve).  (d) Simulated (black squares) longitudinal distribution that removes inter- and intra-cycle interference,  the prediction of GRT (red curve), and that of the classical trajectory Monte Carlo (CTMC) method (blue dashed line). The simulated momentum spectrum is obtained by solving the TDSE of the hydrogen atom. The laser wavelength is 800 nm and the intensity is 87 TW/$cm^{2}$. The black line in panel (c) denotes the position corresponding to 2 times ponderomotive potential ($2U_p$).}
\end{figure}

\paragraph{Numerical simulations.---}
To validate GRT,  we solve the time-dependent Schr\"{o}dinger equation (TDSE) of a hydrogen (H) atom in an LP field  with a generalized pseudo-spectral method\cite{Tong1997}.
We compare the simulated momentum distribution with the prediction of Eq.\ (\ref{7_5_4}) in Fig.\ 2. To apply Eq.\ (\ref{7_5_4}), we need to first determine the GTs according to a different longitudinal momentum $p_z$ by solving differential Eqs.\ (2)-(3) numerically. In panel (a), the GTs of ionized electrons are illustrated corresponding to $p_z=0.2, 0.4$ and $0.8$. It is shown that, because of the long-range Coulomb potential, the emergent impact parameter $b_g$ increases rapidly as $p_z$ decreases.

  Quantitatively, in panel (b), we scale the transverse momentum $p_x$ by the corresponding $1/b_g$ (depending on $p_z$), i.e., with respect to $p_x b_g$ at different $p_z$ from $0.2$ to $0.8$.
  Comparing the normalized simulation results in dots with the black solid curve of $J_0^2$, we clearly find that all the curves collapse onto the  theoretical profile, particularly around the central peak. These calculations confirm that the transverse distribution can be well depicted by Eq.\ (\ref{7_5_4}).

We now focus on the longitudinal momentum distribution. In  Fig.\ 2(c), the distribution calculated from our theoretical result of  Eq.\ (\ref{7_5_4}), i.e., $\varpi P_{\bot g} b_g$, is plotted with a red solid curve. Except for some rapid oscillations, its trend agrees with the numerical result of TDSE in the low-energy regime near the ionization threshold as well as in the high-energy regime exceeding $2U_P$.

The rapid oscillations represent the inter- and intra-cycle interference\cite{arbo}.
 To remove these effects, we take  only  a half-cycle ionization burst\cite{Meckel_2014} in our solving TDSE. We can then observe a smooth longitudinal momentum distribution  as shown by the black squares in panel (d), which perfectly agrees with our theory based on the Glory scattering\cite{Supp}. The result from the classical trajectory Monte Carlo (CTMC) method\cite{book_JieLiu, Hatsagortsyan_2010} is plotted with a blue dashed line for comparison, and a clear quantum enhancement is demonstrated in the regime of small $p_z$\cite{arbo_2008}.

\begin{figure}[t]
\includegraphics[width=1.0\linewidth]{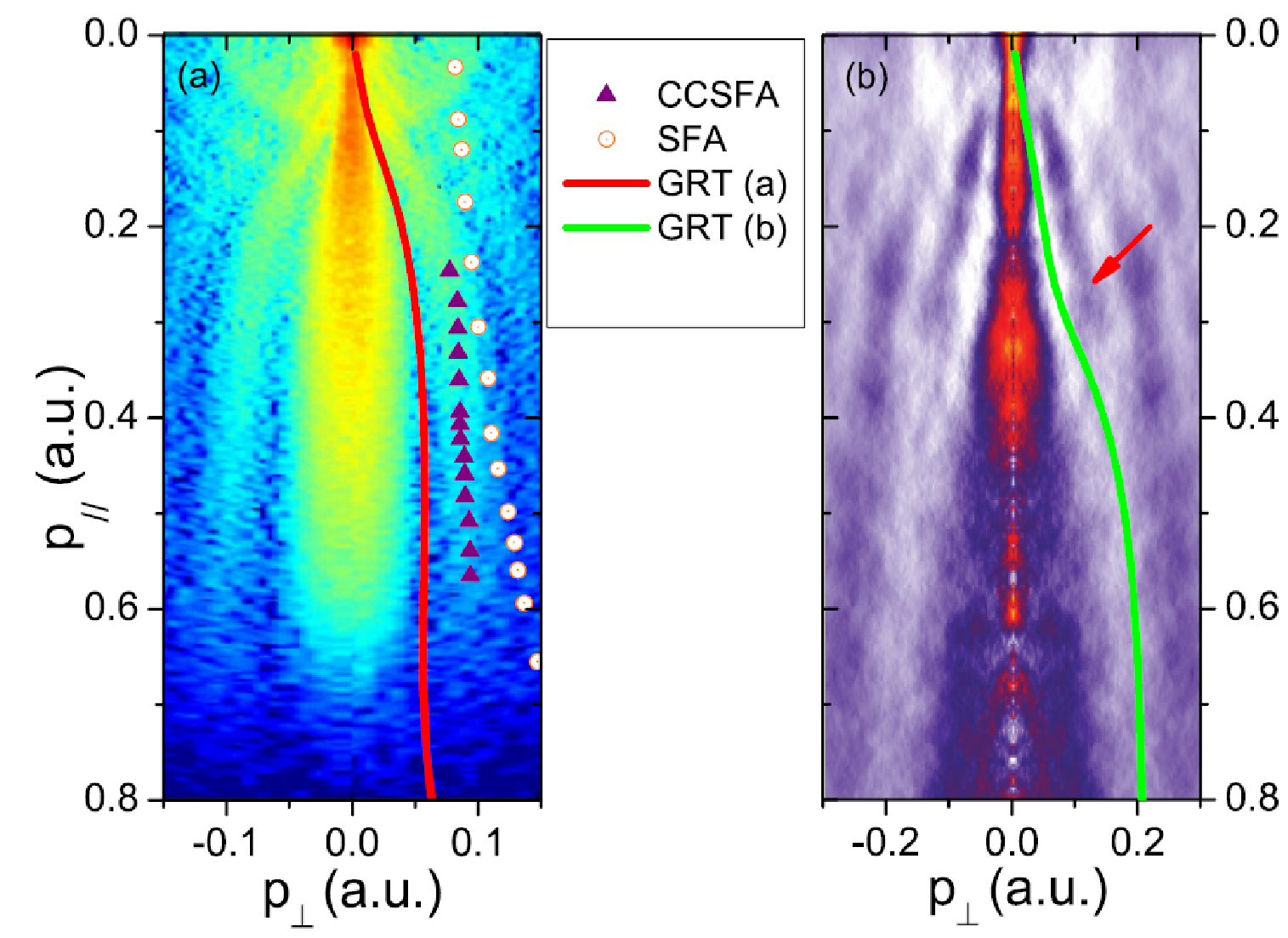}
\caption{(Color online)Experimental holographic pattern and positions of the first dark fringe calculated by SFA (orange open circles), CCSFA (purple solid triangles), and GRT (red solid line in (a) and green in (b)). The experimental data of the metastable (6s) Xe atoms in (a) are extracted from Ref.\ \cite{huismansPRL} as well as SFA and CCSFA results, and that in (b) using argon is extracted from Ref.\ \cite{Hickstein_2012}.   The laser parameters are (a) wavelength of 7000 nm and intensity of $7.1\times10^{11}W/cm^2$, and (b) wavelength of 1300 nm and intensity of $7.5\times10^{13}W/cm^2$.}
\end{figure}

\paragraph{Forward holographic fringe.---}
We now apply GRT to strong-field photoelectron holography, which has been attracting considerable attention since the experiment using the metastable (6s) xenon (Xe) atoms\cite{Huismans, huismansPRL} because it might provide a new approach to ultrafast photoelectron spectroscopy\cite{Bian2012,Hickstein_2012,Meckel_2014,Haertelt2016,Zhouym, HeMR_2018}.
The principle of the holography is to extract the information of electron motion in atoms from the final momentum  spectrum of ionized electrons that exhibit various interference structures. This technique needs a theoretical inverse calculation, and among the many theories, the strong-field approximation (SFA)\cite{KFR}, Coulomb-corrected SFA (CCSFA)\cite{ccsfa} and adiabatic theory of scattering\cite{Zhouym} are commonly utilized, even though some controversies remain long-standing  unresolved.

Figs.\ 3(a) and (b) are two typical holographic patterns that present clear fringe structures. The experiments in these figures are using  7000 nm\cite{huismansPRL} and 1300 nm\cite{Hickstein_2012} laser fields, respectively.
GRT of Eq.\ (\ref{7_5_4})
predicts that  the border of the central brightest lobe of the 2D momentum spectrum, (i.e., the location of the shadow fringe)  should be determined by the first zero point of the Bessel function given by the relation $p_{\bot} \approx 2.4/b_g$, where the emergent impact parameter $b_g$ with respect to $p_\parallel$ of the GTs can readily be calculated from Eqs.\ (2)-(3) according to  the experimental parameters of atoms and lasers. The results are plotted in Figs.\ 3(a) and (b) as a solid red curve and green curve, respectively,  and these curves show a very good agreement with experimental observations. For comparison, in panel (a), we also plot the results predicted by other theories, such as SFA (orange open circles) and CCSFA (purple solid triangles). Both of them evidently deviate from the experimental data: the SFA prediction locates near the experimental secondary dark fringe, whereas the CCSFA prediction shifts to the secondary bright fringe.

In the theoretical framework of SFA or CCSFA, the forward holographic patterns arise from the interference of two semiclassical paths, i.e., the magenta paths in Fig.\ 1(a). The two paths approach the same asymptotic momentum but have a phase difference $\Delta S$. In the treatment of SFA, CCSFA, or some other theories
\cite{Hickstein_2012, Zhouym}, the coherent summation of the two paths leads to  oscillations of type $\cos\Delta S$ that are responsible for the fringe structures in the holograph.

However, the axial caustic singularity leads to the breakdown of the above scenario\cite{book_Nussenzveig, Berry2015}. In the polarization direction, an infinite number of  semiclassical paths can approach the same  final momentum,  and the quantum interference of these trajectories will dominate the holographic fringe structure.
These infinite semiclassical paths  are integrated to give rise to  a pattern of  $\sim J_0^2(l_g \theta)$
according to GRT,
where the angular momentum $l_g=p_z b_g$ and $\theta \sim p_{\bot}/p_z$. Analogous to the  optical diffraction of a ring source \cite{book_Nussenzveig}, here, $p_z=p_\parallel$ and $b_g$ constitute the wave momentum and the radius of the light source, respectively.

Notice that the axial caustic of infinite co-dimension is a stronger singularity than any other singularities such as fold, cusp and butterfly\cite{Goreslavskii1999, Rost_2012, Raz2012}, according to catastrophe theory\cite{Berry1976}. Our GRT also differs from the intuitive model that interprets the holographic pattern as the superposition of plane and spherical  waves of photoelectrons\cite{Hickstein_2012}.

\begin{figure}[t]
\includegraphics[width=1.0\linewidth]{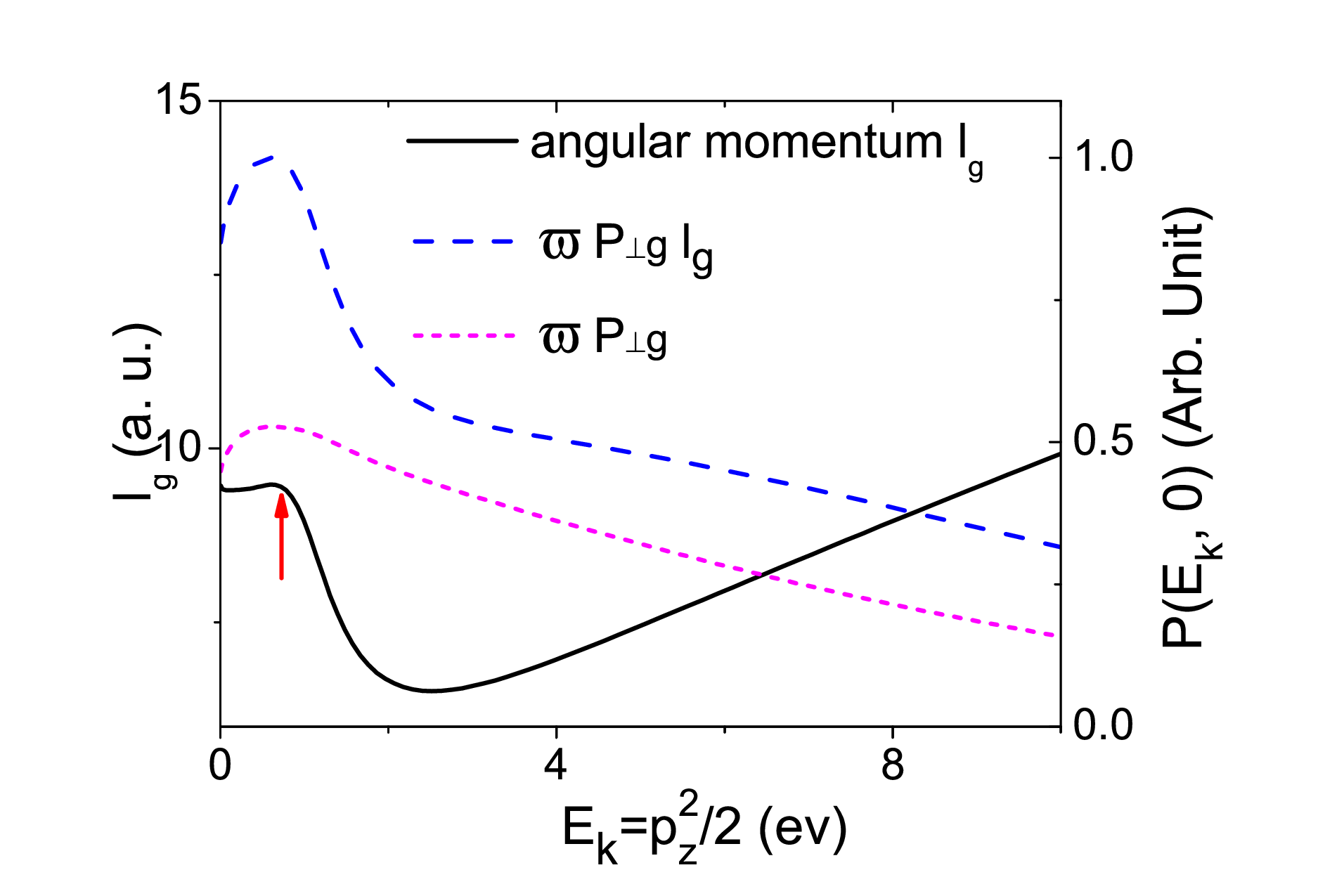}
\caption{(Color online)
$l_g$ of the GT with respect to their asymptotic momentum $p_z$, or  electron energy $E_k=p_z^2/2$, is represented by the black solid line. The angle-resolved photoelectron energy spectrum along the polarization direction, i.e., $P(E_k, \theta=0)\sim\varpi P_{\bot g}l_g$, is plotted with a blue dashed line. For comparison,  the result of $\varpi P_{\bot g}$ is plotted with the magenta short dashed line.
Ar atoms are used. The laser parameters are the same as in Fig.\ 3(b).
}
\end{figure}

 GRT further predicts an abrupt broadening of the holographic central lobe, as indicated  by the red arrow in Fig.\ 3(b).  According to the property of Bessel functions, we have the width of $\delta\theta \approx 2.4/l_g$. This relation indicates that the transition in the holographic structure embodies a sudden change in the angular momentum (i.e., $l_g$) of the GT. In Fig.\ 4, our calculation exhibits a clear sudden decrease (e.g., small peak), labeled by the red arrow, whose location  corresponds to the  sudden increase in the width of the main lobe shown in Fig. 3 (b). By scrutinizing the GTs, we find that the  mechanism behind the sudden change is soft-recollision\cite{Rost_2012}, i.e., the electron revisits its parent ion, with its longitudinal coordinate and momentum  approaching zero simultaneously,  but it avoids head-on recollision by keeping $x\neq0$. In the presence of multiple returns, we expect a sequence of abrupt broadening of the holographic central lobe\cite{Hickstein_2012, Quan_2009}.

 The longitudinal distribution is also dramatically influenced by the GT.
For small $\theta$, the transition amplitude $M_{\vec{p}}$ in Eq.\ (\ref{7_5_4}) can be transformed into the angle-resolved photoelectron energy spectrum $P(E_k, \theta)$ as\cite{Supp}
\begin{eqnarray}
P(E_k, \theta)\sim \varpi P_{\bot g}l_g J_{0}^2(l_g\theta).
\label{7_5_5}
\end{eqnarray}
This result indicates that the distribution of the kinetic energy along the polarization direction is proportional to $\varpi P_{\bot g} l_g$, i.e., $P(E_k, \theta=0)\sim \varpi P_{\bot g} l_g$.
The kinetic energy spectrum is plotted in Fig.\ 4 as a blue dashed curve. For comparison,  the result of $\varpi P_{\bot g}$ is plotted with a short magenta dashed line.
 This result clearly indicates that quantum coherent Glory rescattering is closely related to the low-energy peak structure\cite{Blaga_2009, Quan_2009}.

 In summary, we report the emergence of forward Glory rescattering in laser-assisted photoionization and develop a nonperturbative approach named GRT to address it.
  Our theory largely bridges the long-standing gap between
the quantum interference picture in atomic tunneling ionization and experimental observations. The theoretical framework can be readily extended to molecules with rotational symmetry.
 Hence, our result
 provides a valuable window to probe the atomic and molecular tunneling configuration and holographic interference structure.

\ \

This work is supported by the National Natural Science Foundation of China (Grants No. 11674034,  No. 11775030, No. 11725417 and No. 11447015) and NSAF (No. U1730449).


\begin{thebibliography}{*}
\bibitem{book_Nussenzveig}
H. M. Nussenzveig, \emph{Diffraction Effects in Semiclassical Scattering} (Cambridge University Press, 1992)
\bibitem{Adam2002}
J. A. Adam, Phys. Rep. {\bf 356}, 229 (2002)


\bibitem{wheeler}
K. W. Ford and J. A. Wheeler, Ann. Phys. {\bf 7}, 259 (1959)

\bibitem{Berry2015}
M. V. Berry, Contemp. Phy., {\bf 56}, 2 (2015)

\bibitem{Poelsema1982}
B. Poelsema, S. T. de Zwart, and G. Comsa, Phys. Rev. Lett. {\bf 49}, 578 (1982)
\bibitem{Hussein1984}
M. S. Hussein, A. L\'{e}pine-Szily, M. M. Saad, and A. C. C. Villari, Phys. Rev. Lett. {\bf 52}, 511 (1984)
\bibitem{Ueda1998}
M. Ueda, M. P. Pato, M. S. Hussein, and N. Takigawa, Phys. Rev. Lett. {\bf 81}, 1809 (1998)
\bibitem{Roberts2002}
T. D. Roberts, A. D. Cronin, D. A. Kokorowski, and D. E. Pritchard, Phys. Rev. Lett. {\bf 89}, 200406 (2002)
\bibitem{AngewChem2005}
V. Aquilanti, E. Cornicchi, M. Teixidor, N. Saendig, F. Pirani, and D. Cappelletti, Angew. Chem. Int. Ed. {\bf 44}, 2356 (2005)
\bibitem{Morette1984}
T.-R. Zhang and C. DeWitt-Morette, Phys. Rev. Lett. {\bf 52}, 2313  (1984)

\bibitem{Smirnova2007}
O. Smirnova, A. S. Mouritzen, S. Patchkovskii, and M. Y.
Ivanov, J. Phys. B {\bf 40}, F197 (2007)

\bibitem{Kleinert}
H. Kleinert, \emph{Path Integrals in Quantum Mechanics, Statistics, Polymer Physics, and Financial Markets} (World Scientific, Singapore, 2009).

\bibitem{Berry1976}
M.V. Berry, Advances in Physics, {\bf 25}, 1 (1976).



\bibitem{Huismans}
Y. Huismans \emph{et al.}, Science {\bf 331}, 61 (2011).





\bibitem{huismansPRL}
Y. Huismans \emph{et al.}, Phys. Rev. Lett. {\bf 109}, 013002 (2012).



\bibitem{Hickstein_2012}
D. D. Hickstein \emph{et al.}, Phys. Rev. Lett. {\bf 109}, 073004 (2012)

\bibitem{Blaga_2009}
C. I. Blaga \emph{et al.}, Nature Phys. {\bf 5}, 335 (2009)

\bibitem{Quan_2009}
W. Quan \emph{et al.}, Phys. Rev. Lett., {\bf 103}, 093001 (2009)



\bibitem{book_JieLiu}
J. Liu, \emph{Classical Trajectory Perspective of Atomic Ionization in Strong Laser Fields} (Springer, 2014).
\bibitem{recollision}
P. B. Corkum, Phys. Today {\bf 64}, 36 (2011).


\bibitem{Tau_2017}
J. F. Tao, Q. Z. Xia, J. Cai, L. B. Fu, and J. Liu, Phys. Rev. A {\bf 95}, 011402(R) (2017)

\bibitem{Meckel_2014}
M. Meckel,	A. Staudte,	S. Patchkovskii, D. M. Villeneuve, P. B. Corkum, R. D\"{o}rner, and M. Spanner, Nat. Physics {\bf 10}, 594 (2014)
\bibitem{landau}
L. D. Landau and E. M. Lifshitz, \emph{Quantum Mechanics} (Pergamon, New York, 1977)

\bibitem{Supp}
 See Supplemental Material
for the
details of the derivation of the photoelectron spectrum, and the implementation of the numerical solution of TDSE, which includes
Refs. [10, 21-23, 25-27].
\bibitem{KFR}
L. V. Keldysh, Sov. Phys. JETP {\bf 20}, 1307 (1965); F. H.
M. Faisal, J. Phys. B {\bf 6}, L89 (1973); H. R. Reiss, Phys.
Rev. A {\bf 22}, 1786 (1980).

\bibitem{landau_mechanics}
L. D. Landau and E. M. Lifshitz, \emph{Mechanics} (Oxford, 1960)

\bibitem{Goreslavski_2004}
S. P. Goreslavski, G. G. Paulus, S.V. Popruzhenko, and N. I. Shvetsov-Shilovski, Phys. Rev. Lett. {\bf 93}, 233002  (2004)
\bibitem{Tong1997}
X. M. Tong and S. I. Chu,  Chem. Phys. {\bf 217}, 119 (1997)

\bibitem{ADK}
M. V. Ammosov,  N. B. Delone and  V. P. Krainov, Zh. Eksp. Teor. Fiz. {\bf 91}, 2008 (1986) [Sov. Phys. JETP {\bf 64}, 1191 (1986)]
\bibitem{ccsfa}
T.-M. Yan, S. V. Popruzhenko, M. J. J. Vrakking, and D. Bauer, Phys. Rev. Lett. {\bf 105}, 253002 (2010)

\bibitem{LiM_2016}
M. Li, J.-W. Geng, M. Han, M.-M. Liu, L.-Y. Peng, Q. Gong, and
Y. Liu, Phys. Rev. A {\bf 93}, 013402 (2016)


\bibitem{arbo}
D. G. Arb\'{o}, K. L. Ishikawa, K. Schiessl, E. Persson, and J. Burgd\"{o}rfer,
Phys. Rev. A {\bf 81}, 021403(R)  (2010)


\bibitem{Hatsagortsyan_2010}
C. Liu and K. Z. Hatsagortsyan, Phys. Rev. Lett. {\bf 105}, 113003 (2010)


\bibitem{arbo_2008}
D. G. Arb\'{o}, J. E. Miraglia, M. S. Gravielle, K. Schiessl, E. Persson, and J. Burgd\"{o}rfer, Phys. Rev. A {\bf 77}, 013401 (2008)

\bibitem{Bian2012}
X.-B. Bian and A. D. Bandrauk,
Phys. Rev. Lett. {\bf 108}, 263003 (2012)


\bibitem{Haertelt2016}
M. Haertelt, X.-B. Bian, M. Spanner, A. Staudte, and P. B. Corkum,
Phys. Rev. Lett. {\bf 116}, 133001 (2016)

\bibitem{Zhouym}
Y. Zhou, O. I. Tolstikhin, and T. Morishita, Phys. Rev. Lett. {\bf 116}, 173001 (2016)


\bibitem{HeMR_2018}
M. He, Y. Li, Y. Zhou, M. Li, W. Cao, and P. Lu,
Phys. Rev. Lett. {\bf 120}, 133204 (2018)














\bibitem{Rost_2012}
A. K\"{a}stner, U. Saalmann, and J.-M. Rost, Phys. Rev. Lett.
{\bf 108}, 033201 (2012).

\bibitem{Goreslavskii1999}
S. P. Goreslavskii and S. V. Popruzhenko, J. Phys. B {\bf 32}, L531 (1999).

\bibitem{Raz2012}
O. Raz, O. Pedatzur, B. D. Bruner and N. Dudovich, Nat. Photon {\bf 6}, 170 (2012).


\end{thebibliography}
\end{document}